\documentclass[preprint,12pt,authoryear]{elsarticle}

\usepackage{times}
\usepackage{graphicx}
\usepackage{epsf}
\usepackage{epsfig}
\usepackage{amsmath}
\usepackage{amssymb}
\usepackage{hhline,multirow}
\usepackage{subfig}
\usepackage{ifpdf}
\usepackage[latin1]{inputenc}
\usepackage{url}

\def\ie{\emph{i.e.}}
\def\eg{\emph{e.g.}}

\newtheorem{algorithm}{Algorithm}

\begin{document}
\title{Semi-fragile watermarking of remote sensing images using DWT, vector quantization and automatic tiling }

\author{Jordi Serra-Ruiz and David Meg\'ias\\
Estudis d'Informàtica, Multimèdia i Telecomunicacions\\
Universitat Oberta de Catalunya \\
Rambla del Poblenou 156, 08018 Barcelona, Spain \\
jserrai@uoc.edu, dmegias@uoc.edu \\
}

\begin{abstract}
A semi-fragile watermarking scheme for multiple band images is presented in this article. We propose to embed a mark into remote sensing images applying a tree structured vector quantization approach to the pixel signatures, instead of processing each band separately. The signature of the multispectral or hyperspectral image is used to embed the mark in it order to detect any significant modification of the original image. The image is segmented into three-dimensional blocks and a tree structured vector quantizer is built for each block. These trees are manipulated using an iterative algorithm until the resulting block satisfies a required criterion which establishes the embedded mark. The method is shown to be able to preserve the mark under lossy compression (above a given threshold) but, at the same time, it detects possibly forged blocks and their position in the whole image.
\end{abstract}


\maketitle

\section{Introduction}
\label{sect:introduction}

Remote sensing images have incremented interest by the research community in the last years, since the news applications of this particular images are often reported. Looking for water in remote planets (recent NASA missions), water pollution control, high precision farming, among others, are well-known uses of remote sensing images~\citep{atkinson99,wong03AE}.
The acquisition of these images require expensive equipment, like an aircraft or satellites. Therefore their economic value must be preserved when a third party pays for these images.

The last new techniques for representation, storage and distribution of digital information have been developed in the last years. Format like MP3 or JPEG2000, and P2P networks for sharing these digital contents, increase the distribution of all kind of files.

This situation has raised the problem of managing authorized and unauthorized copying, illegal distribution through the Internet and manipulation of digital information. To prevent illegal copies and the alteration of digital files (audio or image), some methods and techniques have been developed to embed a watermark into the digital files. This watermark must be imperceptible and can used to determine the integrity of the digital files. In this authentication process, two different approaches, namely semi-fragile watermarking and robust watermarking, can be used.
The modification and free distribution without permission of the digital contents are become easier in the last years, therefore the research in content protection~\citep{Yu01} and distribution of these has raised. A control modification system is needed to protect these contents.

Both fragile and semi-fragile watermarking schemes can be used for tampering detection and localization. In fragile schemes, all modifications of the content are detected as tampering. Therefore any kind of lossy compression or filters can not be applied to the marked image without removing the embedded mark. On the other hand, semi-fragile schemes allow some degree of compression and small modifications of the marked images. This allows, for example, to create a compressed version of the image which can be distributed electronically (possibly with a reduced price) but maintaining the original watermark embedded in content. Hence, semi-fragile watermarking makes unnecessary the marking of different versions of the same image independently and, thus, reduces the cost required to distribute different versions of the same image with different degrees of quality. If the client has access to a compressed version of the image, he or she may check the integrity of the image (discarding tampering). After that, he or she may be interested in getting access to the original uncompressed image at a higher price. For example, the schemes ~\citep{Yeung97,Fridrich02-SecureFragileAuthentication} embed a watermark into an image in such a way that the embedded information is destroyed or modified if the image is tampered.
It is modified or destroyed when the marked image is manipulated.

Robust watermarking methods are those for which the mark is detected after strong alterations. Robust watermarking allows different types of manipulations, including compression, filtering or geometrical distortions. Usually, robustness requires reducing the transparency of the embedded mark, \ie{} the marked file is significantly distorted. In this case, a trade-off between robustness and perceptual quality must be achieved. Robust watermarking schemes have proven successful in order to protect images in several ways, such as resolving authoring disputes or detecting changes in the images aimed to produce a forged copy, as shown in~\cite{lin00detection} and~\cite{minguillon03RS}.

In remote sensing imaging applications, the most useful schemes aimed to detect changes in the image are semi-fragile watermarking systems. Semi-fragile schemes are able to overcome some minor modifications, as those produced by near-lossless compression, but reveal the existence of manipulations, such as copy-and-replace ``attack'', or excessive information removal by means of cropping or lossy compression. Most existing semi-fragile watermarking applications can be applied on monochromatic images, and thus, can be easily extended to multispectral or hyperspectral images (remote sensing images) by processing the multiple bands of these images independently.

It is worth pointing out the difference between watermarking and hashing schemes. The watermarking process alters the original data file by modifying the content in order to embed the mark. Most schemes require a common (secret) key both at the embedder at the detector which is usually introduced to endorse the system with security features. The detection process only needs the (secret) key to determine whether the image has been altered and where. With a blind detection method, no further information is needed to detect the mark (\ie{} the original unmarked content is not required). On the other hand, hashing methods generate a given number (\eg{} a checksum) which is needed in the detection process. If tamper localization is required, the hashing values of different areas of the image must be generated and checked at the receiver side. Hence, the detector needs all the hashing values related to the original image. In addition, if multiple versions of the image are created (\eg{} with different compression ratios) multiple hashing values would be required (a set of values for each different version). Consequently, the semi-fragile watermarking approach makes it possible to protect the content of different versions of a hyperspectral image using a single procedure. Furthermore, the watermarking scheme can be used even for different hyperspectral images with the same embedding key. This is not possible with hashing techniques, which require the computation of a different set of hashing values for each different version of each image.

There are some previous works dealing with satellite image watermarking~\citep{Qin04wavelet,Ho05pinned,Wang05,Caldelli06,Sal08,Tamhankar03}.
~\cite{Ho05pinned} use a satellite image and decompose it into two mutually orthogonal sub-fields, but only uses one band of the satellite image and only one field for watermarking purposes. ~\cite{Qin04wavelet} present a semi-fragile watermarking scheme based on wavelet transforms. The edge and texture of the remote sensing image are extracted and the watermark is embedded only in the edge character. In this case, once again, only one band is marked. ~\cite{Caldelli06} present a scheme to embed an authentication mark using the method described by~\cite{Fridrich02-SecureFragileAuthentication} to mark and compress the image at the same time. This scheme uses only one band to embed the mark into each block to detect manipulations. ~\cite{Wang05} present a watermarking scheme to preserve a digital content, but only uses one band of the hyperspectral image of the Indian Remote Sensing System. ~\cite{Sal08} describe an evolutionary algorithm which marks an image based on the manipulation of the discrete cosine transform (DCT) computed for each band of the image. Finally, ~\cite{Tamhankar03} describe a method to embed one mark into the hyperspectral image using the whole signature, but it does not allow compression of the hyperspectral image. This adaptive watermarking method based on the redundant discrete wavelet transform (RDWT) is a fragile scheme.

In this paper, a semi-fragile watermarking scheme specifically developed for remote sensing images is presented. This method works with all the bands at the same time and provides uniform protection for multispectral and hyperspectral imaging applications. The method can be tuned to embed the mark according to band relevance, depending on the content and the signatures (also known as the spectral reflectance curve) to be protected.

Usually, fragile or semi-fragile watermarking schemes of multiple band images also consider only one band or process each band separately ~\citep{ekici04JEI,Caldelli06}. It is possible to work with images using only one band for computing where and how to embed the watermark, but then, when multiband images are marked with the same method, the bands are usually marked separately or only one band or a subset of bands are marked. Note that, if the bands are marked separately, the changes in the signature curves can be uneven (some values can be increased and others decreased, for the same pixel).
Hence, the shapes of the signatures may vary, which may lead to a misclassification of the image (for example, a different material could be identified in the image). Because of this, a method which preserves the shapes of the signatures is highly demanded, and this can be achieved by working with the signatures as a whole.

The method suggested in this paper uses the hyperspectral image as a whole applying a vector quantization approach. The image is segmented in three-di\-men\-sion\-al blocks of a variable size which determines the spatial resolution of the embedding and detection algorithm. For each block, a tree with an endmember (real values read by the remote sensor for each pixel region) of the remote sensing image is built and these endmembers are manipulated by applying the wavelet transform, for each block, in order to increase the robustness against possible near-lossless compression attacks. Finally, the block is manipulated using an iterative algorithm until the resulting block (TSVQ tree) satisfies some criterion. The image is modified according to a secret key which produces a different criterion for each block in order to avoid copy-and-replace attacks. This key determines the internal structure of the tree and also the resulting distortion.

This paper is organized as follows. In Section~2, the coding of remote sensing images by means of tree structured vector quantization is reviewed. In Section~3, the watermarking strategy and the mark embedding and mark retrieval processes are described. Section~4 presents the results obtained with the suggested scheme for the chosen experimental corpus, and analyses the basic parameters that determine the results. Finally, the most relevant conclusions of this work are drawn in Section~5.

\section{Background}
\label{sect:background}

In this section, an overview of three basic concepts used in the proposed method in this paper, namely remote sensing images, lossy compression and vector quantization, are briefly introduced here.

\begin{figure}[ht!]
 \center
  \leavevmode
  \ifpdf  \includegraphics[width=0.8\textwidth]{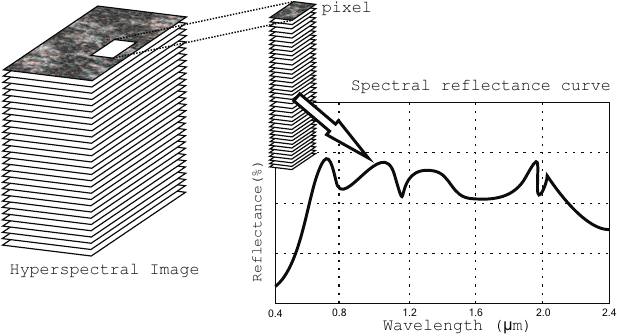}
  \else   \includegraphics[width=0.8\textwidth]{grafico2.eps}
  \fi
  \caption{Example of the signature curve for a pixel.}\label{fig:capes_signatura}
\end{figure}

\subsection{Remote sensing images}
\label{subsect:remote_sensing}
Remote sensing images store information about a broad area of the surface of the Earth.  Each pixel of these images is represented by a set of values, named signature, obtained by a special sensor for different frequencies of the light spectrum (bands). The signature of each pixel of the remote sensing image is related to the different materials which can be found in that area, such as water, forest or minerals.

The construction scheme of this information is shown in Figure~\ref{fig:capes_signatura}.

\begin{figure}[ht!]
    \leavevmode
    \ifpdf  \includegraphics[width=0.8\textwidth]{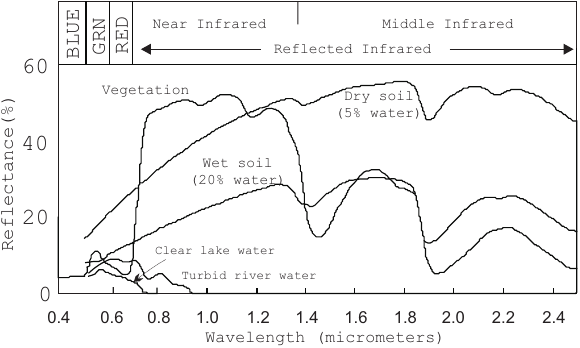}
    \else   \includegraphics[width=0.8\textwidth]{grafico1.eps}
    \fi
  \caption{Sample signatures for different materials.}\label{fig:signatures1}
\end{figure}

Figure~\ref{fig:signatures1} shows the different signatures for a light reflectance for clear lake water, turbid river water, vegetation, dry soil and wet soil, as presented in~\cite{signatures}.

One of the most relevant problems to handle these images is their huge size. A typical hyperspectral image covering a small region of a few kilometres contains millions of pixels, and each pixel is represented by several bands, depending on the sensor type. As an example, the Airbone Visible / Infrared Imaging Spectrometer (AVIRIS)~\citep{aviris} images contain 224 bands and, usually, 16 bits are used for the values in each band. Images with lower resolution are often referred to as ``multispectral". This is the case for Landsat images~\citep{landsat}, which use 8 bands for each pixel and 8 bits for the values of each band. The number of bands in a remote sensing image determines their name, that is, multi, hyper or ultraspectral.

One of the techniques to reduce the large amount of data storage is to apply a lossy compression method~\citep{Aiazzi06,Mielikainen06}. ``Lossy'', means that, once the input image has been encoded and later decoded, the recovered image is not exactly the same as the original input image, but some information has been removed. Although some specific applications do not allow any kind of information removal, some information losses are allowed in many other situations. However, the noise introduced by the lossy compression process must be kept below a given threshold to avoid damaging relevant information.

\subsection{Lossy compression of remote sensing images}
\label{subsect:lossy}

Lossy compression methods remove information which is not significant for image reconstruction. This is the key issue in lossy image compression, because the
information removed should depend on the user purposes. Several criteria for
image quality can be defined, as described in~\cite{cristophe05GRS}, depending on the desired goal. Preliminary experiments~\citep{minguillon00-1,minguillon00-2} show that it is possible to achieve relatively high compression ratios without removing critical information.

The general coding scheme must be adapted to the particular characteristics of
the source, in order to maximize both the compression ratio and the image
fidelity. In this case, the most important issue is that the remote sensing
images are 3D, where two dimensions are spatial but the third one is spectral.
An ideal compression method would take advantage of this fact, trying to
exploit both spatial and spectral redundancy. Coding each band separately is
suboptimal as spectral redundancy is not exploited, while applying 3D coding
schemes does not take into account the difference between spatial and spectral
dimensions. Several authors have proposed 3D transformations to de-correlate
spatial and spectral redundancy. For example, \cite{motta05} present a survey on hyperspectral image compression. In this paper, vector quantization is used for lossy compression porpoise, and all bands are processed at the same time.

\subsection{Discrete Wavelet Transform}
\label{subsect:DWT}

\subsection{Watermarking}
\label{subsect:water}
Watermarking~\citep{Peti99book} consists of imperceptibly embedding some information into a cover object (\eg{} a remote sensing image) to produce a marked version of the same object. The watermarking process alters the original data file by modifying the content in order to embed the mark. Most schemes require a common (secret) key both at the embedder at the detector which is usually introduced to endorse the system with security features. The detection process needs the (secret) key to determine whether the mark is or not embedded in the test object. With a blind detection method the original unmarked content is not required.

Semi-fragile schemes are able to overcome some minor modifications, as those produced by near-lossless compression, but reveal the existence of strong manipulations (or ``attacks'' in the watermarking literature), such as copy-and-replace or excessive information removal by means of cropping or lossy compression.  On the other hand, robust watermarking allows stronger types of manipulations, including compression, filtering or geometrical attacks. Usually, robustness requires significant distortion in the marked image and a trade-off between robustness and perceptual quality must be achieved. Robust watermarking schemes have proven successful in order to protect images in several ways, such as resolving authoring disputes or detecting changes in the images aimed to produce a forged copy, as shown in~\cite{lin00detection,minguillon03RS}.

In remote sensing imaging applications, the most useful schemes aimed to detect changes in the image are semi-fragile watermarking systems. Most existing semi-fragile watermarking applications can be applied on monochromatic images, and thus, can be easily extended to multispectral or hyperspectral images (remote sensing images) by processing the multiple bands of these images independently.

\subsection{Vector Quantization and Tree Structure Vector Quantization}
\label{subsect:vq}
Vector Quantization (VQ), as described in~\cite{gersho92}, makes it possible to compress an image in an optimal manner from the Shannon's rate-distortion theory point of view. As detailed in \citep{gersho92,gray98}, \cite{shannon48} showed that, given a coding rate, the least distortion achievable by vector quantisers of any kind is equal to a function, subsequently called the Shannon distortion-rate function, which is determined by the statistics of the source and the measure of distortion. Shannon's rate-distortion theory establishes the minimal amount of information which must be communicated over a channel so that the source can be approximately reconstructed at the receiver without exceeding a given distortion. Thus, applying VQ compression systems, the resulting image minimizes (locally) distortion for a given compression ratio. However, VQ compression can be computationally prohibitive. Hence alternative methods known to be suboptimal need to be explored for practical applications.

Tree Structured Vector Quantiser (TSVQ) is a suboptimal strategy which works
starting with an initial centroid as the codebook, that is, a tree with a
single leaf, and then a quality criterion is applied (Mean Square Error). If
there is room for improvement, the leaf with higher distortion is split into
two similar centroids, and then the Linde-Buzo-Gray algorithm,
described in~\citep{linde80}, is applied for computing the new centroids. This
process is repeated until a general quality criterion is achieved or when all
leaves contain only equal samples within the same leaf, meaning that a perfect
tree has been built. For large images, when the number of training vectors
is also large, the resulting tree is usually quite deep, although it might be
very unbalanced. Nevertheless, the number of possible subtrees is large enough
to explore the possibilities of finding feasible subtrees for watermark
embedding.

Finally, the original image is coded using the selected subtree, replacing each original vector by the closest centroid, that is, by the centroid representing
all the elements in the leaf where the original vector falls. This selection is performed starting from the root of the tree, and choosing the closest centroid until a leaf is reached. Unlike the generation of the initial tree $T$, coding the tree \emph{selected }is a very fast operation which can be performed very efficiently.

\section{Semi-fragile watermarking scheme}
\label{sect:scheme}

The watermarking scheme presented in this paper is described in the following
sections. In Subsection~\ref{sect:embedding}, the embedding method based on tiling, DWT and TSVQ is deteiled, while Subsection~\ref{sect:detection} describes the detection process and how it can be applied to locate forged regions.

\subsection{Mark Embedding Process}
\label{sect:embedding}

\begin{figure}[ht!]
  \begin{center}
    \leavevmode
    \ifpdf  \includegraphics[width=0.97\textwidth]{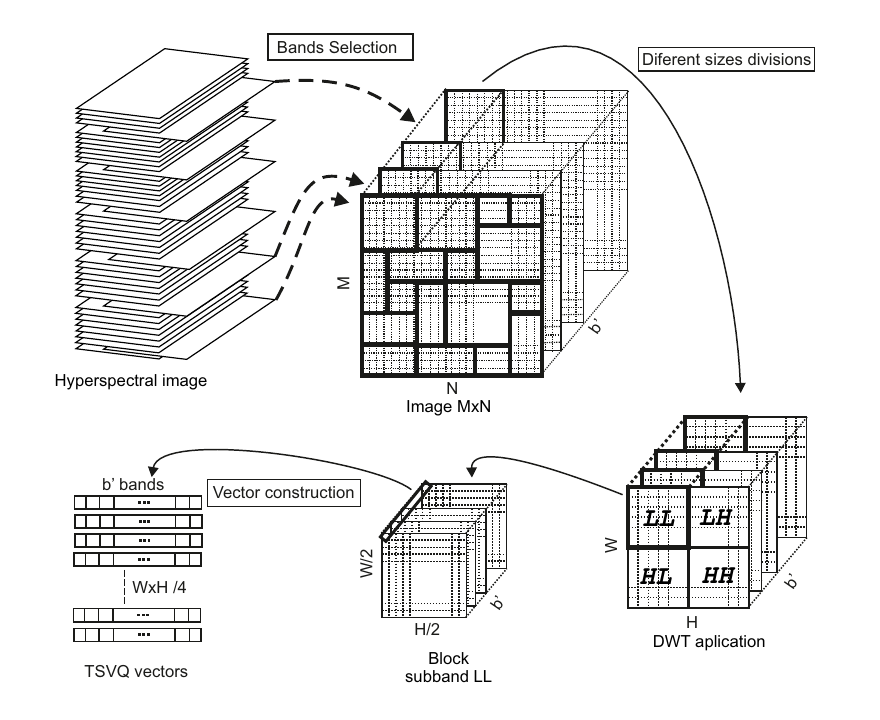}
    \else   \includegraphics[width=0.97\textwidth]{Dibujo_TI_WT.eps}
    \fi
    \caption{Generation of the signature vectors for TSVQ.}
    \label{fig:signatures}
  \end{center}
\end{figure}

Let us consider an original three-dimensional hyperspectral image $I$ of size
$M \times N \times b$ samples, where $b$ stands for the number of bands. The image $I$ is segmented
in different blocks or tiles $W_i \times H_i \times b'$, where $W_i\leq M$ and $H_i\leq N$ are the size in pixels of each block (different for each block) and $b'\leq b$ is the number of selected bands. This block division makes it possible to detect specific tampered regions in the attacked images. Using blocks of different sizes makes the scheme against some attacks based on analysing the properties of the blocks by an attacker who may know the watermarking process. If the attacker tries to recreate the properties of a forged are he or she must know the exact tiling of the image. Keeping the tiling secret is thus, an additional measure of security.

As detailed in Fig.~\ref{fig:signatures}, the original $512 \times 512$ image is thus divided into smaller blocks of different sizes which are marked separately. The specific tiling (the size and position of each block) for each image must be recorded and transmitted to the detector as part of the secret key of the scheme. Although choosing a manual tiling of the image is possible, this paper proposes an automatic tiling process in order to obtain more homogeneous regions. The automatic tiling process is described below. This process is run for all the selected bands $b_k$. The selected band is divided into mini-blocks of size $32 \times 32$ samples. These blocks are compared with their neighbours in order to build bigger blocks of the following possible sizes $W_i,H_i \in \{32, 64, 96, 128\}$, hence the smallest possible block size is $32 \times 32$ and the biggest possible block size is $128 \times 128$. The unassigned blocks of size $32 \times 32$ samples are grouped in biggest contiguous super-blocks of $128 \times 128$ samples (or $4 \times 4$ sub-blocks). For each super-block, the average of the samples of its sixteen sub-blocks $B_{i,j}$ with $i,j\in\{1,2,3,4\}$ is obtained and compared with that of a possible block $B'$ of size $32j \times 32i$ containing $B_{i,j}$.

\begin{figure}[ht!]
  \begin{center}
    \leavevmode
    \ifpdf  \includegraphics [width=0.4\textwidth] {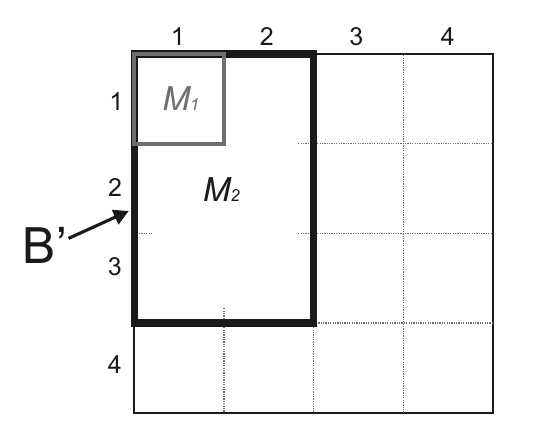}
    \else   \includegraphics [width=0.4\textwidth] {dise_bloques.eps}
    \fi
    \caption{Candidate block $B'$ including sub-block $B_{3,2}$.}
    \label{fig:grupstiling2}
  \end{center}
\end{figure}

Fig.~\ref{fig:grupstiling2} illustrates how the average of the samples of the sub-block $B_{3,2}$ is compared with a that of a possible block $B'$ of size  $64\times 96$ samples. If the difference between these averages is lower than a given threshold $l$, then the candidate $B'$ is selected as a possible block. After the process is completed, the biggest possible block $B'$ fulfilling this condition is selected. More precisely, the averages $M_1$ and $M_2$ are obtained as follows:

\begin{equation*}
\label{eq:blocs2}
\begin{array}{l}
     M_1= \frac{\displaystyle\sum_{n=1}^{32}\sum_{m=1}^{32}B_{i,j}(n,m)}{32 \times 32}, \\
     \\
     M_2= \frac{\displaystyle\sum_{n=1}^{32i}\sum_{m=1}^{32j}B'(n,m)}{32i \times 32j}, \\
\end{array}
\end{equation*}

\noindent where $i$ and $j$ are the coordinates of the sub-block $B_{i,j}$ being analysed, $B_{i,j}(\cdot,\cdot)$ are the samples of the mini-block and $B'(\cdot,\cdot)$ are the samples of the candidate block. The block $B'$ is selected as eligible if it turns out that $|M_2-M_1| \leq l$.  Finally, the biggest eligible block $B'$ is selected and included in the tiling for the chosen band.

After this process, $b'$ different tilings for the $b'$ selected bands are available. Finally, the tiling producing the greatest number of blocks (the smallest homogenous blocks) is chosen for the whole image and kept as part of the secret key of the method.

\begin{figure}[ht!]
  \begin{center}
    \leavevmode
    \ifpdf  \includegraphics [width=0.8\textwidth] {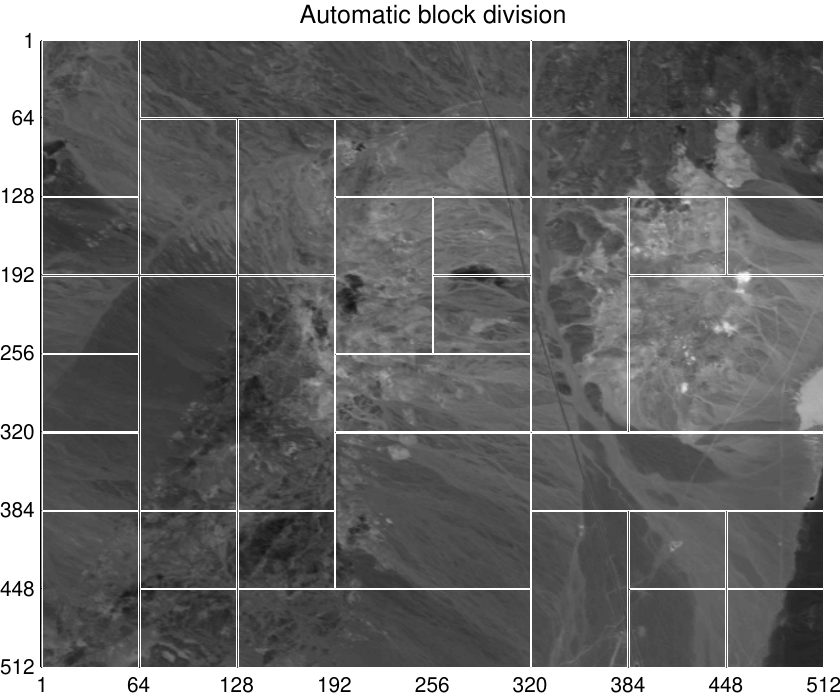}
    \else   \includegraphics [width=0.8\textwidth] {Auto_Blocs.eps}
    \fi
    \caption{Block division obtained after applying the tiling process.}
    \label{fig:grupstiling_generica_reals}
  \end{center}
\end{figure}

Fig.~\ref{fig:grupstiling_generica_reals} shows and example of automatic block division using this method. As it can be seen, generally, there are bigger blocks for homogenous regions and smaller blocks for more variable areas. Bigger blocks are more convenient to reduce computations and achieve a more even distortion, whereas smaller blocks provide specific protection for variable areas.

\begin{algorithm}[Tiling process]
\label{alg:tiling}
~
\begin{enumerate}
    \item for all the band of the whole image do
    \begin{enumerate}
        \item split the band in $32\times32$ samples blocks.
        \item calculate the average value $M_1$ of each sub-blocks $B_{i,j}$~(fig.~\ref{fig:grupstiling2}).
        \item for all the $B_{i,j}$ ~do
            \begin{enumerate}
            \item calculate the 16 average values $M_{2,k}$ of ~$4\times4$ sub-blocks. (where $k = 1\ldots16$)
            \item compare $M_1$ and $M_{2,k}$ averages. (where $k = 1\ldots16$)
            \item select the biggest sub-block $M_{2,k}$ that the average difference is lower than $l$ (threshold).
            \end{enumerate}
    \end{enumerate}
    \item select the final distribution of block from the band that has more number of blocs.
\end{enumerate}
\end{algorithm}


After the tiling process, in order to achieve robustness about near-lossless compression, the discrete Wavelet transform (DWT) is applied for each block and TSVQ vectors are constructed
using the $LL$ sub-band of the DWT.  The vectors formed with the coefficients of the $LL$ sub-band of each block are replaced by
similar values obtained from the leaves of a TSVQ tree, with minimal
distortion and enforcing a particular property (e.g. the entropy) which will be checked at the
detector side. Then, the BFOS algorithm~\cite{breiman84} prunes the generated
tree with the selected criterion to obtain all the subtrees in the compression
ratio-distortion curve. A parameter of the resulting TSVQ tree determines the
subtree which is required to obtain the specific compression of the image
block, for example the entropy. It is worth pointing out that the
compression ratio does not take into account the individual bands but the
signatures as a whole (in fact the selected bands $b'$ within the signature), since we are using a vector quantization approach.
Hence, this proposal is different from other semi-fragile watermarking methods
in the literature which process each band separately. Fig.~\ref{fig:signatures} summarizes the embedding process.

A different criterion (marking property) is selected for each of the tiles using a Pseudo-Random Number Generator (PRNG), the seed of which is also required in the detection process. Thus, this seed is also part of the secret key (together with the specific tiling of the image). The $LL$ sub-band of each block of the original image is compressed with a different compression ratio, according to the selected TSVQ criterion. This new $LL^*$ sub-band is processed with the TSVQ process again, using the same parameters as in the first iteration, and this process is repeated until the generated
block satisfies the selected property, producing a modified $LL'$ sub-band which is used in the marked image.

As manipulations to the marked image are concerned, a large modification in any
single band or a small number of bands is considered unacceptable in remote
sensing images, because these attacks introduce an uneven change in the
spectral signature. Only slight distortions affecting the whole signature are
accepted, such as near lossless compression, but only up to a certain degree.

To detect possible tampering attacks, such as copy-and-replace, or pasting one part of another image into the marked image, the PRNG sequence is used to determine the values of the property used as a criterion to select the
compression subtree in the pruning algorithm. Thus, it is much more difficult
to find a pattern revealing the watermarking scheme properties, reducing the
possibilities of manipulating an image or modifying it using another region of
the same image. Note that copy-and-replace attacks are even more difficult to carry out with the non-uniform tiling of the image, since it is very unlikely that an attacker may guess the exact size and location of a particular block to replace it by a different one (and with the same marking property).


\begin{figure}[ht]
  \begin{center}
    \leavevmode
    \ifpdf  \includegraphics[width=0.98\columnwidth]{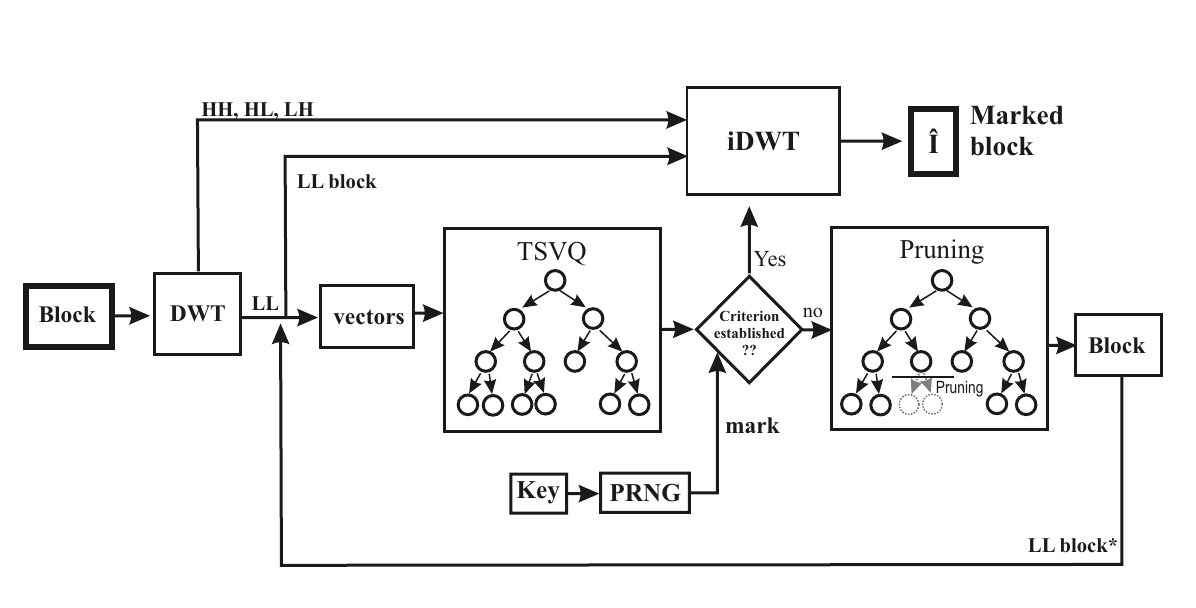}
    \else   \includegraphics[width=0.98\columnwidth]{schememark_WT.eps}
    \fi
    \caption{Block diagram of the embedding method.}
    \label{fig:scheme}
  \end{center}
\end{figure}

In the suggested scheme, shown in Fig.~\ref{fig:scheme}, the entropy is the chosen parameter to select which
particular tree is used to build the marked $LL'$ sub-band. For each block, the
pseudo-random sequence determines the entropy of the compression tree. Finally,
the BFOS algorithm generates a table with all the possible subtrees which are
in the convex hull, minimizing the distortion for a given compression ratio.
Usually, both compression ratio and MSE are used for generating the convex
hull, but the BFOS algorithm may be applied with any other criteria which might be
more suitable for watermarking purposes or for joint \cite{caldelli04IGARSS} compression and
watermarking. Once a specific subtree is
selected to generate the modified $LL$ sub-band, the resulting marked content is
obtained with its centroids yielding a specific value for the entropy which represents the embedded mark. The result of the TSVQ process, the $LL'$ DWT sub-band obtained with the
centroids, is then combined with the original $LH$, $HL$ and $HH$ DWT sub-bands (kept in the
first step) of the block and the inverse DWT is applied to generate
the marked block. Finally, all blocks are joined to construct the marked
image.

As already remarked, the secret key is the concatenation of the tiling information with the seed of the PRNG, the quantization step chosen of the entropy and the initial (reference) entropy value. In any case, the size of the secret key is relatively small (typically less than 2 or 3 kilobytes) and should be transmitted using a secured channel.  The embedding process is summarised in Algorithm \ref{alg:embedding}.

\begin{algorithm}[Embedding process]
\label{alg:embedding}
~
\begin{enumerate}
  \setcounter{enumi}{-1}
  \item Pre-process in order to suppress the noise of the sensors.\footnote{This (optional) step may be achieved by slightly compressing-decompressing the image with some codec, such as JPEG2000, e.g. using the KaKaDu software of \cite{kakadu}.}
  \item \label{ti-bands} Select the $b'$ bands to be embedded.
  \item Run the tiling process of Algorithm \ref{alg:tiling} and save the positions and sizes of the selected blocks and build the image blocks as shown in Fig.~\ref{fig:signatures}. \label{ti-tiling}
  \item Choose the property to use for embedding (e.g. entropy), its initial (reference) value and the quantization step.
  \item Choose the seed of the PRNG and initialize it. Keep the seed as part of the secret key.
  \item For all the blocks of the image selected in Step \ref{ti-tiling} do
  \begin{enumerate}
  \item \label{ti-wt} Apply the DWT and save the $LH$, $HL$, and $HH$ sub-bands.
    \item \label{ti-prng} Generate a new number using the PRNG and check that it is not used in a previous block (otherwise generate a new one until no already used number results).
    \item \label{ti-quant} Choose the value of the marking property (e.g. entropy) according to the pseudo-random number obtained in the previous step (\ref{ti-prng}).
    \item \label{ti-select} Start the TSVQ pruning process and choose a subtree satisfying the property obtained in the previous step (\ref{ti-quant}).
    \item \label{ti-generate} Generate a new sub-band $LL^*$ using the subtree obtained in the previous step (\ref{ti-select}).
    \item Check if the property chosen in Step \ref{ti-quant} is satisfied with the tree constructed with $LL^*$ sub-band obtained in Step \ref{ti-generate}. If this property is not fulfilled, go  back to Step \ref{ti-select} using the new $LL^*$ sub-band. If the property is achieved, let $LL':=LL^*$.
  \item Apply the inverse integer DWT to the resulting $LL'$ sub-band together with the original $LH$, $HL$ and $HH$ ones stored in Step \ref{ti-wt}.
  \end{enumerate}
  \item Build the final marked image by joining all the marked blocks (according to the saved sizes and positions) and the bands which were not selected in Step \ref{ti-bands}).
\end{enumerate}
\end{algorithm}

\subsection{Tampering Detection and Localization}
\label{sect:detection}

\begin{figure}[ht]
  \begin{center}
    \leavevmode
    \ifpdf  \includegraphics[width=0.75\columnwidth] {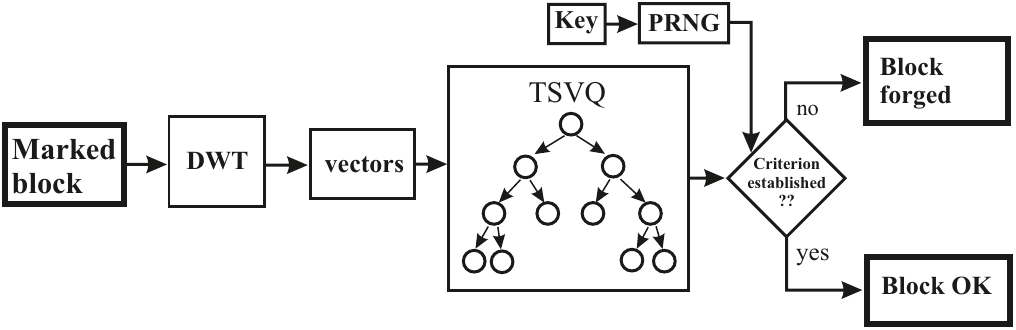}
    \else   \includegraphics[width=0.75\columnwidth] {schemedetect_WT.eps}
    \fi
     \caption{Block diagram of the detection method.}
    \label{fig:schemedetect}
  \end{center}
\end{figure}

The mark detection process for a (possibly forged) image is analogous to the
mark embedding process and it only requires the secret key. The detection method is blind since it does not need the original image. The scheme is depicted in
Fig.~\ref{fig:schemedetect}. First of all, the
same bands must be extracted from the test image and the specific tiling is reproduced using the secret key. Then the integer DWT must be applied to each band of each block. The detection of a modification of the image can be performed by
checking if the same criterion used in the mark embedding scheme is satisfied.
If the tree constructed with the $LL$ DWT sub-band of each block verifies the
criterion given by the pseudo-random sequence generated with the seed contained in secret key, then the block is assumed to be authenticated. Otherwise, the block is detected as forged. Therefore, with this
method it is possible to detect and locate tampered regions in a marked image. The resolution of the tampering detection varies from $32\times 32$ to $128\times 128$ pixels (in steps of 32 pixels), which correspond to the block sizes provided by the tiling process.

\begin{algorithm}[Tampering detection and localization]
\label{alg:tamperdetect}
~
\begin{enumerate}
\item Retrieve the $b'$ marked bands of the test image (this information is contained in the secret key).
\item Retrieve the image tiling: block sizes and positions (this information is also stored in the secret key).  \label{td-blocks}
\item Retrieve the property used for embedding (e.g. entropy), its initial (reference) value and the quantization step (also from the secret key).
\item Retrieve the seed of the PRNG from the secret key and initialize it.
\item For all the blocks built in Step \ref{td-blocks} do
\begin{enumerate}
\item Apply the DWT to obtain the sub-bands $LL$, $LH$, $HL$ and $HH$ for that block.
\item Generate a new number using the PRNG and check that it is not used in a previous block (otherwise generate a new one until no already used number results).
\item Choose the value of the marking property (e.g. entropy) according to the pseudo-random number obtained in the previous step.
\item Build the TSVQ tree with the $LL$ sub-band and verify whether it satisfies the property computed in the previous step. If the value of the property is not verified, report this block as forged. Otherwise, the block is authenticated.
\end{enumerate}
\end{enumerate}
\end{algorithm}

\begin{table*}[ht]
\begin{center}
\caption{PSNR of the marked bands (using the Daubechies db1 integer DWT).} \label{tab:PSNR}
\begin{tabular}{||c|c|c|c||c|c|c|c||}
  \hhline{|t:====:t:====:t|}
  Band & PSNR & PMP & Adi & Band & PSNR & PMP & Adi \\
  \hhline{|:====::====:|}
  1 & 60.97 & 32.72 & 18.45 & 9 & 57.94 & 32.94 & 24.80 \\
  \hhline{||-|-|-|-||-|-|-|-||}
  2 & 61.30 & 32.48 & 16.75 & 10 & 62.78 & 32.50 & 15.11 \\
  \hhline{||-|-|-|-||-|-|-|-||}
  3 & 61.98 & 32.53 & 16.03 & 11 & 62.70 & 32.54 & 15.36 \\
  \hhline{||-|-|-|-||-|-|-|-||}
  4 & 62.01 & 32.61 & 16.47 & 12 & 61.08 & 32.70 & 18.79 \\
  \hhline{||-|-|-|-||-|-|-|-||}
  5 & 62.14 & 32.44 & 16.04 & 13 & 61.62 & 32.68 & 17.27 \\
  \hhline{||-|-|-|-||-|-|-|-||}
  6 & 62.18 & 32.52 & 16.06 & 14 & 59.86 & 32.69 & 19.89 \\
  \hhline{||-|-|-|-||-|-|-|-||}
  7 & 62.53 & 32.50 & 15.07 & 15 & 60.65 & 32.75 & 18.96 \\
  \hhline{||-|-|-|-||-|-|-|-||}
  8 & 62.08 & 32.54 & 15.84 & 16 & 58.87 & 32.87 & 23.95  \\
  \hhline{|b:========:b|}
\end{tabular}
\end{center}
\end{table*}


\begin{figure}[ht]
  \begin{center}
    \leavevmode
    \subfloat[]{
    \ifpdf  \includegraphics[width=0.4\textwidth] {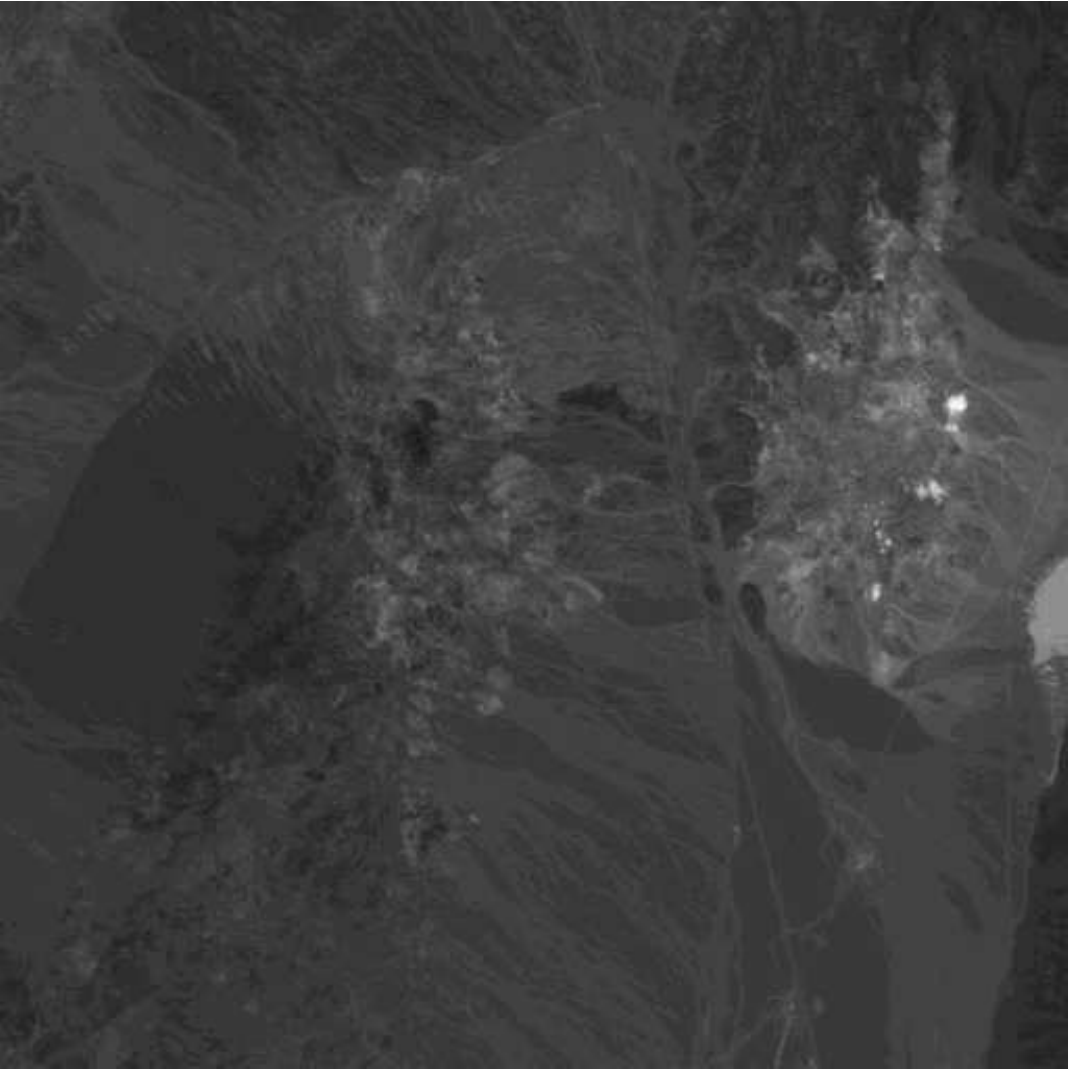}
    \else   \includegraphics[width=0.4\textwidth] {banda2.eps}
    \fi\label{suba}}
    \subfloat[]{
    \ifpdf  \includegraphics[width=0.4\textwidth] {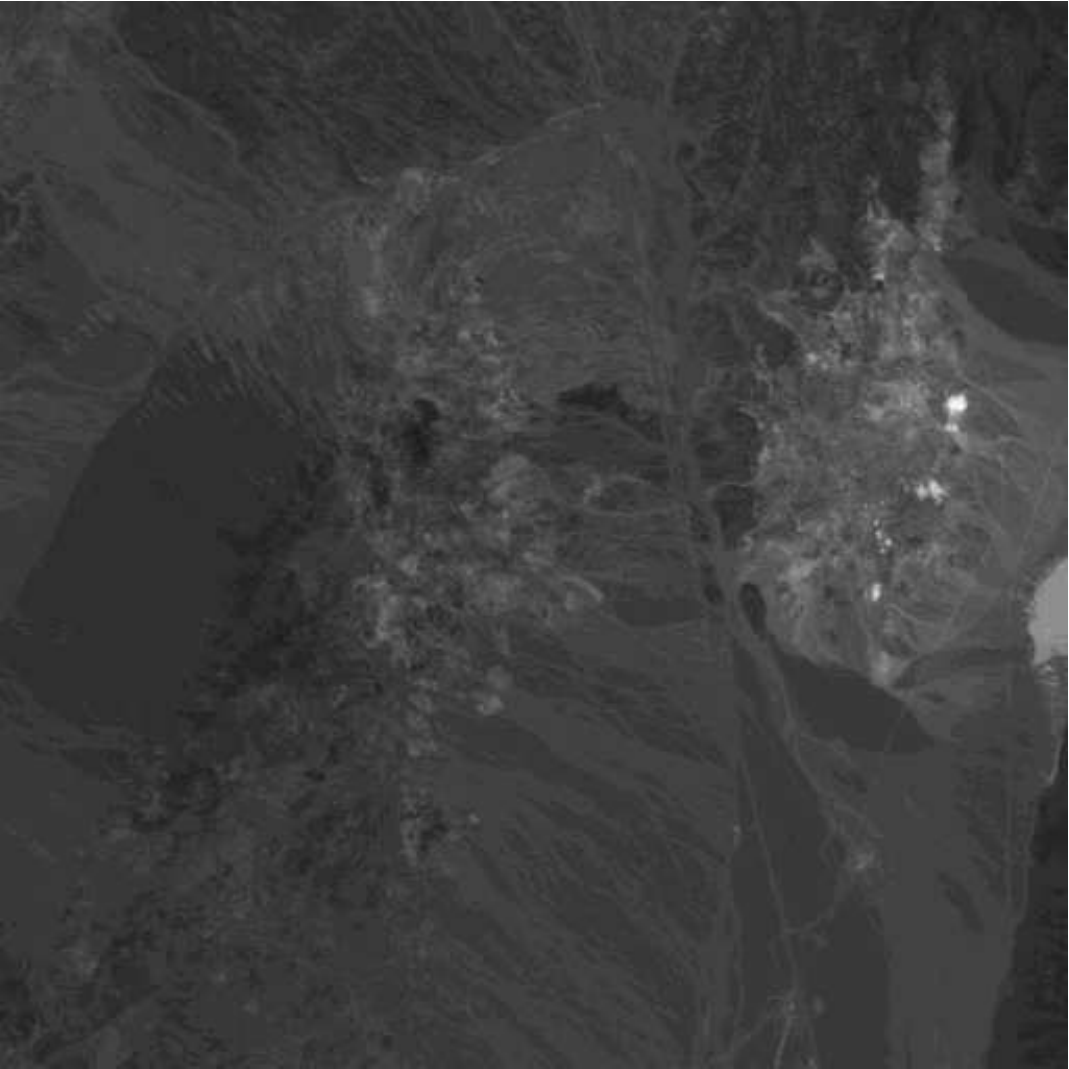}
    \else   \includegraphics[width=0.4\textwidth] {banda_mark2.eps}
    \fi\label{subb}}
    \caption{Original (a) and marked (b) images for band 2.}
    \label{fig:imagemarked}
  \end{center}
\end{figure}

\section*{Acknowledgements and disclaimer}

{\small This work is partially supported by the Spanish Ministry of Science and
Innovation and the FEDER funds under the grants TSI2007-65406-C03-03 E-AEGIS
and CONSOLIDER-INGENIO 2010 CSD2007-00004 ARES.}

\bibliographystyle{elsarticle-harv}
\bibliography{article.bbl}
\end{document}